\documentclass[onecolumn,
 amsmath,amssymb,
 aps,
]{revtex4-1}

\usepackage{graphicx}
\usepackage{dcolumn}
\usepackage{xcolor}
\usepackage[section]{placeins}

\usepackage[hyperfootnotes=true]{hyperref}
\hypersetup{colorlinks, citecolor=blue, linkcolor=red, urlcolor=purple}

\begin{document}

\preprint{APS/123-QED}

\title{On the jets ejected after the inertial collapse of cavities} 

\author{Jos\'e M. Gordillo}
\email{jgordill@us.es}
\author{Francisco J. Blanco--Rodr\'{i}guez}%
\affiliation{\'Area de Mec\'anica de Fluidos, Departamento de Ingenier\'ia Aeroespacial y Mec\'anica de Fluidos, Universidad de Sevilla, Avenida de los Descubrimientos s/n 41092, Sevilla, Spain
}%

\date{\today}

\begin{abstract}
Motivated by the results in Gordillo and Blanco-Rodriguez, 'Bubble bursting jets are driven by the purely inertial collapse of gas cavities', \emph{Phys. Rev. Lett., Submitted} (2023) \cite{PRL2023}, where it is found that bubble bursting jets are driven by a purely inertial mechanism, here we present a study on the dynamics of the jets produced by the collapse of gas cavities of generic shape when the implosion is forced by a far field boundary condition expressing that the flow rate per unit length, $q_\infty$, remains constant in time. Making use of theory and of numerical simulations, we first analyze the case of a conical bubble with a half-opening angle $\beta$ when the value of $q_\infty$ is fixed to a constant, finding that this type of jets converge towards a purely inertial $\beta$-dependent self-similar solution of the equations in which the jet width and velocity are respectively given, in the limit $\beta\ll 1$, by $r_{jet}\approx 2.25\tan\beta\sqrt{q_\infty\tau}$ and $v_{jet}\approx 3 q_\infty/(2\tan\beta\sqrt{q_\infty\tau})$ respectively, with $\tau$ indicating the dimensionless time after the jet is ejected. For the case of parabolic cavities with a dimensionless radius of curvature at the plane of symmetry $r_c$ our theory predicts that $r_{jet}\propto \left(2 r_c\right)^{-1/2}\left(q_\infty \tau\right)^{3/4}$ and $v_{jet}\propto q_\infty \left(2r_c\right)^{1/2} \left(q_\infty\tau\right)^{-3/4}$, a result which is also in good agreement with numerical simulations. The present results might find applications in the description of the very fast jets, with velocities reaching up to $1000$ m s$^{-1}$, produced after a bubble cavitates very close to a wall and in the quantification of the so-called bazooka effect. 
\end{abstract}

\maketitle

\section{Introduction}

It is known that the type of high speed liquid jets produced after the implosion of a bubble, widely used in applications related with the cleaning of surfaces \citep{APLOhl}, is partly responsible for the structural damage in cavitating flows \citep{PRFLechner} and also plays a key role in the dispersion of contaminants \citep{Bouriba,PRLBird}. These high speed jets also possess applications in medicine, where they could be used in needle-free drug injection systems \citep{Tagawa} and are commonly observed after the impact of either a drop \citep{Michon,Thoroddsen_2018,Thoroddsen_2020,JFM2021} or a solid \citep{PRL2009,JFM2010} against a gas-liquid interface. Up to our knowledge, and in spite of the number of recent contributions on the subject, there does not exist a commonly accepted framework capable of predicting the dynamics of the myriad of types of fast and thin inertial liquid jets produced after the implosion of a cavity. Hence, it will be our main purpose here to present a theory to describe the dynamics of the jets ejected after the collapse of slender cavities whose predictions will be compared with the results of numerical simulations carried out using \texttt{GERRIS} \citep{Popinet2003,Popinet2009}.

The velocities of the jets produced after the implosion of a cavity depend very much on the initial geometry of the interface from which they are ejected, which justifies to classify the jets in, at least, two different categories: i) the jets produced at the axis of an initially spherical surface and ii) the jets produced at the base of a truncated conical surface. For instance, the type of jets reported by \cite{Antkowiak,IvoJFM}, which can be analyzed as a function of the initial near-axis non uniform radial velocity field at the interface \citep{JFMImpulsive} and are employed, for instance, in ink-jet printing applications \citep{HajimePRApplied,SegersPRApplied} belong to the type of jets emerging from the bottom of a spherical cap. The jets produced by the cavitation of bubbles near boundaries \citep{Blake,Supponen}, also belong to the same type of jets when the so-called stand-off parameter, which expresses the ratio between the initial distance of the bubble center to the boundary and the maximum bubble radius, is of order unity or larger. Indeed, when the value of the stand-off parameter is $\lesssim O(0.1)$, \cite{PRFLechner} presented numerical results retaining compressibility effects in the simulations revealing that the jets produced by cavitation bubbles near a rigid boundary do not emerge from an initially smooth, locally spherical cap, but from the base of a truncated conical interface. This numerical result was later on confirmed experimentally by \cite{Reuter}, who measured velocities of the order of $\sim 100$ $m/s$ for the case of jets produced by cavitation bubbles emanating from a spherical cap i.e., for values of the stand-off parameter of order unity, and much larger velocities, of the order of $\sim 1000$ $m/s$, for the case of jets emerging from the base of a truncated conical bubble, a result in agreement with the original finding reported by \cite{PRFLechner}. Moreover, \cite{Reuter} realized that the jets emanating from the base of the locally conical bubble are the result of a boundary-parallel cylindrical flow converging towards the axis of symmetry and, motivated by this fact, also reported the results of simplified numerical simulations neglecting compressibility effects, obtaining results in agreement with their own experimental measurements. 

It will be the main purpose in this contribution to describe both numerically and theoretically the so-called `conical jets' namely, the jets emerging from the base of a truncated conical interface. The conical jets, which are comparatively faster than those emerging from a spherical cap, are not only found in the collapse of cavitation bubbles for small values of the stand-off parameter, but also in other types of natural flows, like in the bursting of bubbles at an interface \citep{Duchemin,Deike,JFM2019,JFM2021,PRL2023} or after the impact of a drop on a liquid pool \citep{Thoroddsen_2018,Michon,Thoroddsen_2020,JFM2021}. Indeed, in these two latter cases, before the jet is ejected, capillary waves traveling along the interface towards the bottom of the cavity deform the initially spherical bubble into a truncated cone \cite{JFM2019}. Moreover, our analysis on the collapse of conical cavities is also extended here to the case of cavities with a generic shape, paying special attention to parabolic cavities, because this is the type of void geometry found when a solid impacts a free surface \citep{PRL2009,JFM2010}.

The paper is structured as follows: in \S \ref{sec:Theory} we present a theory describing the ejection of jets from slender cavities whose collapse is forced by a constant value of the flow rate per unit length and, in this section, we also compare the predictions of our theory with the results of numerical simulations described in Appendix \ref{appA}. Conclusions are presented in \S \ref{sec:Con}.

\section{A theory on the ejection of jets from collapsing cavities}
\label{sec:Theory}

Here we consider an axisymmetric bubble of characteristic length $L_c$ filled with a gas of density $\rho_g$ and viscosity $\mu_g$ which collapses driven by a liquid of density $\rho$, viscosity $\mu$ and interfacial tension coefficient $\sigma$ flowing towards the axis of symmetry with a characteristic velocity $V_c$. Using $L_c$, $L_c/V_c$ and $\rho V^2_c$ as the characteristic scales of length, time and pressure, the physical situation under study here can be described in terms of the following dimensionless parameters:
\begin{equation}
We=\frac{\rho\,V^2_c\,L_c}{\sigma}\, ,\quad Re=\frac{\rho\,V_c L_c}{\mu}\, ,\quad m=\frac{\mu_g}{\mu}\, ,\quad \Lambda=\frac{\rho_g}{\rho}\, .\label{WeRe}
\end{equation}

\begin{figure*}
	\includegraphics[width=1\textwidth]{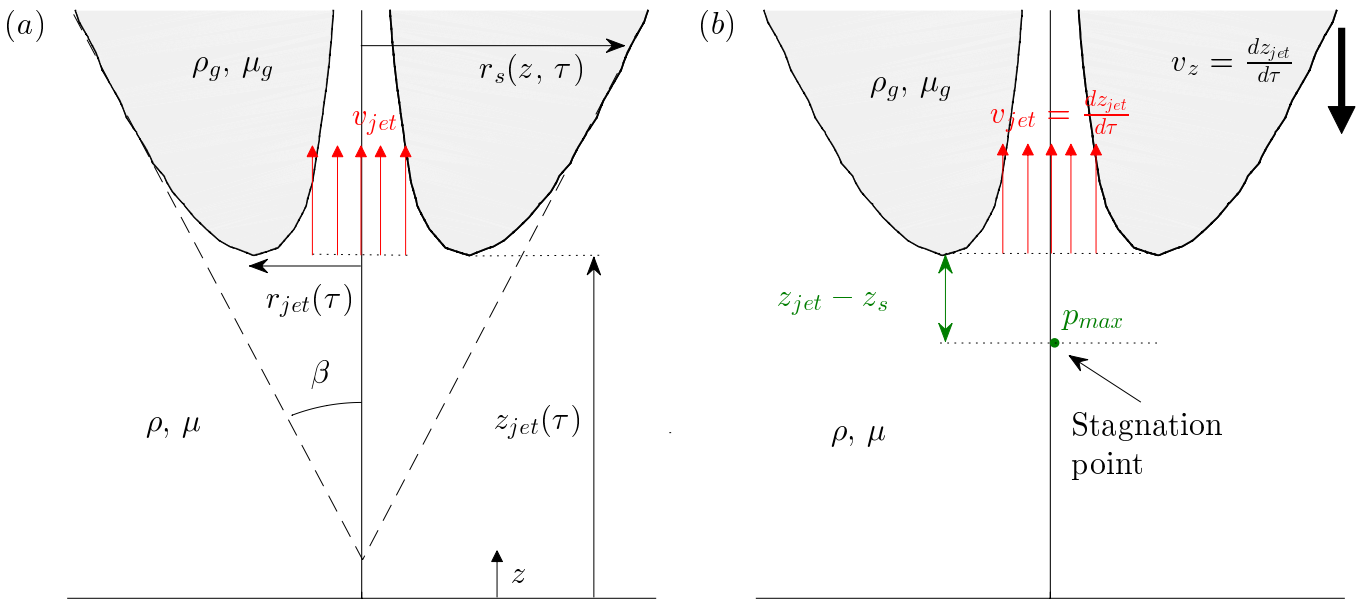}
	\caption{(a) Sketch illustrating the definitions of the time-dependent variables used to analyze the jet ejection process: the interface is located at $r=r_s(z,\tau)$, the base of the jet is the point at the interface of coordinates $r=r_{jet}(\tau)$, $z=z_{jet}(\tau)$ where  $\partial r_s/\partial z=0$ and the jet velocity is defined as $v_{jet}(\tau)=v_z(r=0,z=z_{jet}+\alpha r_{jet})$ with $\alpha\sim 1$, being $z=z_{jet}(\tau)+\alpha r_{jet}(\tau)$ the vertical coordinate within the jet where the pressure approximately relaxes to that of the gas. (b) Sketch of the flow in a frame of reference moving vertically with the velocity $d z_{jet}/d\tau$. In this frame of reference, there exists a stagnation point of the flow located at a distance from the base of the jet $z_{jet}-z_s\propto r_{jet}(\tau)$ where the pressure is maximum and is equal to $p_{max}(\tau)$.} 
   \label{fig1}
\end{figure*}

From now on, dimensionless variables will be written using lower case letters to differentiate them from their dimensional counterparts, written in capitals, $r$ and $z$ will respectively indicate the dimensionless radial and axial spatial coordinates in a cylindrical coordinate system and $\tau$ will indicate the dimensionless time after the jet is ejected. In \cite{PRL2023}, $L_c=R_b$, with $R_b$ indicating the initial radius of the spherical bubble and $V_c=\sqrt{\sigma/(\rho R_b)}$ is the so called capillary velocity. Therefore, for the case of bubble bursting jets considered in \cite{PRL2023}, $We=1$ and $Re=Oh^{-1}$, with $Oh=\mu/\sqrt{\rho R_b \sigma}$ the so called Ohnesorge number, which must be such that $Oh\ll 1$ for a jet to be ejected from the bottom of the collapsing bubble  \cite{JFM2020}. The results in \cite{PRL2023} show the relevance played by the two time-dependent functions $z_{jet}(\tau)$ and $r_{jet}(\tau)$ characterizing the position of the interface of equation $r-r_s(z,\tau)=0$ in the description of the jet dynamics, see Fig. \ref{fig1}, so it will be our main purpose in this contribution to find analytical expressions for both $r_{jet}(\tau)$ and $z_{jet}(\tau)$ in the limit $\tau\ll 1$ of interest here.

Since the results in \cite{PRL2023} indicate that $r_{jet}\propto \sqrt{q_\infty\tau}$ and $v_{jet}\propto \sqrt{q_\infty/\tau}$ when the collapse of a conical cavity is forced by a far field boundary condition expressing that $q_\infty$ remains constant in time, the value of the local Reynolds number characterizing the flow at the base of the jet, $Re_l=Oh^{-1}v_{jet} r_{jet}\propto Oh^{-1}$, will remain constant in time  and the value of the local Weber number will diverge as $We_l=v^2_{jet} r_{jet}\propto \tau^{-1/2}\gg 1$ in the limit $\tau\ll 1$ of interest here \cite{PRL2023}. Then, provided that the Reynolds number at the scale of the bubble is large i.e., provided that $Oh^{-1}\gg 1$ for the case considered in \cite{PRL2023}, both capillary and viscous effects can be neglected, an approximation which is further supported by the numerical results presented in Appendix \ref{appA}. Moreover, since we will not consider the effects of the gas on the jet ejection process, the solutions to be deduced next will not depend on any of the dimensionless parameters defined in Eq. (\ref{WeRe}); however, as it will be shown below, our results will strongly depend on the value of the dimensionless flow rate per unit length $q_\infty$ and also on the initial geometry of the cavity i.e., on the function $r_s(z,\tau=0)$.

It was explained in \cite{PRL2023} that, in physical terms, the reason why the radial flow rate per unit length directed towards the axis of symmetry remains approximately constant in time is due to the fact that liquid inertia ensures that this quantity does not change appreciably during the very short instants of time following the emergence of the jet \cite{JFM2010,JFM2021,PRL2023}.  Then, during the very short transient $\tau\ll 1$ during which the conical cavity collapses, the flow rate $q_\infty$ acts as the far field boundary condition driving the jet ejection process \cite{PRL2023}. The constancy of the flow rate per unit length forces the inward motion of the conical cavity walls and, therefore, the jet is issued as a mere consequence of mass conservation \cite{JFM2019,JFM2021,PRL2023}. 

\begin{figure*}
	\includegraphics[width=\textwidth]{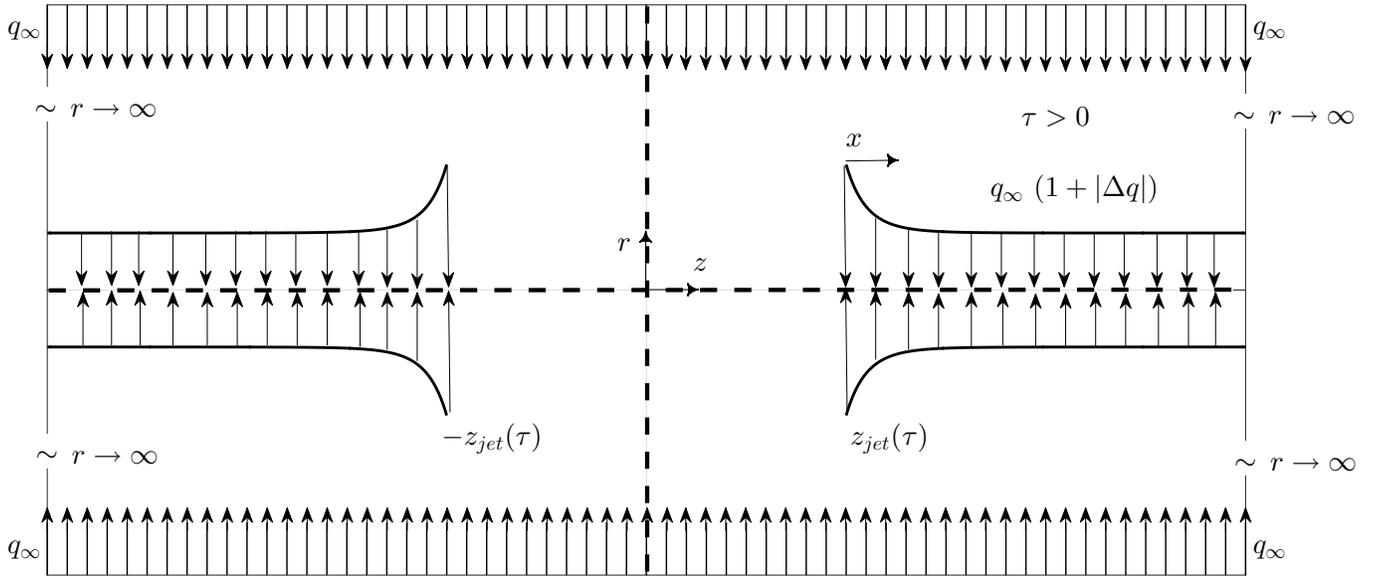}
	\caption{The sketch illustrates that, for instants of time $\tau\ll 1$, the irrotational velocity field satisfying the boundary conditions in Eqs. (\ref{farfield})-(\ref{cc2}) is a line of sinks of intensity $q_\infty\left(1+|\Delta\,q(z,\tau)|\right)$ extending along the axis in the region $|z|>z_{jet}(\tau)$ with $q_\infty \Delta\,q(z,\tau)$ such that $\partial\phi/\partial z=0$ at $r=r_s(z,\tau)\simeq 0$ since $r_s\ll 1$, see Eqs. (\ref{cc3})-(\ref{Deltaq1}).} 
   \label{fig2}
\end{figure*}

Consequently, since in view of the discussion in the paragraphs above, viscous and capillary effects can be neglected, we consider here the case of an incompressible flow in which vorticity is zero, enabling us to express the velocity field in terms of a velocity potential $\phi$ i.e., $\mathbf{v}=\nabla\phi$, with $\phi$ satisfying the Laplace equation $\nabla^2\phi=0$. We then analyze, at short times, $\tau\ll 1$, the implosion of an axisymmetric cavity, symmetric with respect to the plane $z=0$ which collapses forced by a far field boundary condition of the type illustrated in Fig \ref{fig2},
\begin{equation}
r\,\frac{\partial\phi}{\partial r}=-q_\infty\quad\mathrm{for}\quad r\rightarrow\infty\, .\label{farfield}
\end{equation}

We also consider that the cavity is slender i.e., the position of the interface satisfies $r_s(z,\tau)\ll 1$ and $\partial r_s/\partial z\ll 1$. Before the jet is ejected namely, for instants of time $\tau<0$, the irrotational velocity field satisfying the boundary condition (\ref{farfield}) is the one generated by a line of sinks of intensity $q_\infty$ located at the axis of symmetry. Due to the fact that the cavity is slender, the velocity field produced by the line of sinks at the axis is, in a first approximation, normal to the interface and, therefore, the potential at $\tau=0^+$, which is when the jet is firstly issued, can be set to $\phi=const$ at $r=r_s(z,\tau=0^+)$. As it is sketched in Fig \ref{fig2}, once the two symmetrical jets are ejected from $z=\pm z_{jet}(\tau)$ for instants of time $\tau>0$, the free interface disappears along the region $|z|< z_{jet}(\tau)$ and two symmetric jets are ejected from $z=\pm z_{jet}(\tau)$. Consequently, 
\begin{equation}
r\frac{\partial\phi}{\partial r}(r=0,z,\tau)=0\quad\mathrm{for}\quad |z|< z_{jet}(\tau)\quad\mathrm{and}\quad \tau>0\label{cc2}
\end{equation}
since, otherwise, the liquid velocity would tend to infinity at the axis. Hence, the irrotational and incompressible velocity field satisfying the boundary conditions (\ref{farfield})-(\ref{cc2}) is the one induced by a line of sinks with intensities $q_{\infty}\left(1+|\Delta\,q(z,\tau)|\right)$ extending along $|z|\geq z_{jet}(\tau)$, see Fig \ref{fig2}. In order to determine the unknown function $\Delta\,q(z,\tau)$ we describe the 'impact' of an axisymmetric flow over $r=0$ and make use of Wagner's theoretical framework \citep{wagner1932}, which was originally conceived to quantify the impact of two-dimensional solids with small deadrise angles over free surfaces \citep{AnnRevKorobkin,howison}. Indeed, notice that the integration in time of the Euler-Bernoulli equation (\ref{E-B}) permits to conclude that, for $\tau\ll 1$, $\phi\simeq const$ at $r=r_s(z,\tau)\ll 1$ for $|z|\geq z_{jet}(\tau)$ because, at $\tau=0^+$ $\phi=const$ at the free interface and $\tau|\nabla\phi|^2/2\rightarrow 0$ as $\tau\rightarrow 0$. Moreover, since $r_s\ll 1$, this boundary condition for the Laplace equation can be linearized by retaining only the first term in the Taylor series expansion of $\phi$ around $r=0$, which yields the following boundary condition for $\partial\phi/\partial z$ at $r=0$:
\begin{equation}
\phi(r=0,z,\tau)=const\Rightarrow \frac{\partial\phi}{\partial z}(r=0,z,\tau)=0\quad \mathrm{for}\quad |z|\geq z_{jet}(\tau)\, ,\label{cc3}
\end{equation}
a condition also expressing that, in the limit of slender cavities of interest here, velocities are normal to the interface.
Now, notice that the solution of the Laplace equation $\nabla^2 \phi=0$ subjected to the boundary conditions (\ref{farfield})-(\ref{cc3}) is a distribution of sinks with intensity $q_\infty$ located at the axis of symmetry and extending along $|z|<\infty$, plus a line of sources of intensity $q_\infty$ that extends along the axis in the region $|z|< z_{jet}(\tau)$, plus a line of sinks with a strength $q_\infty\Delta\,q(z,\tau)$ that extends along the axis in the region $|z|\geq z_{jet}(\tau)$. Notice that the distribution $q_\infty\Delta\,q(z,\tau)$ is needed in order to preserve the flow rate per unit length imposed by the far field boundary condition (\ref{farfield}) and to compensate the flow rate which is not suctioned through the axis along the region $|z|<z_{jet}$. Hence, $\Delta\,q(z,\tau)$ is determined from the solution of the integral equation that results from imposing the boundary condition expressed by Eq. (\ref{cc3}):
\begin{equation}
\begin{split}
&\int_{-z_{jet}}^{z_{jet}}\frac{\left(z-z_0\right)d\,z_0}{\left[\left(z-z_0\right)^2\right]^{3/2}}+\int_{z_{jet}}^\infty\frac{\Delta\,q(z_0)\left(z-z_0\right)d\,z_0}{\left[\left(z-z_0\right)^2\right]^{3/2}}+\int_{-\infty}^{-z_{jet}}\frac{\Delta\,q(-z_0)\left(z-z_0\right)d\,z_0}{\left[\left(z-z_0\right)^2\right]^{3/2}}=0\, .
\end{split}\label{Deltaq0b}
\end{equation}
The first integral in Eq. (\ref{Deltaq0b}) corresponds to the axial velocity generated at $r=0$ and $z>z_{jet}(\tau)$ by the line of sources of intensity $q_\infty$ located at $|z|<z_{jet}$, whereas the second and third integrals correspond to the axial velocities at $r=0$ generated by the distribution of sinks of intensity $q_\infty\Delta\,q(z,\tau)$ located at $|z|\geq z_{jet}$, see the sketch in Fig \ref{fig2}. Notice that the contribution of the third integral is rather small when compared with that of the second integral in Eq. (\ref{Deltaq0b}), this being the reason why this term will be neglected in what follows.

\begin{figure*}
	\centering
	\includegraphics[width=\textwidth]{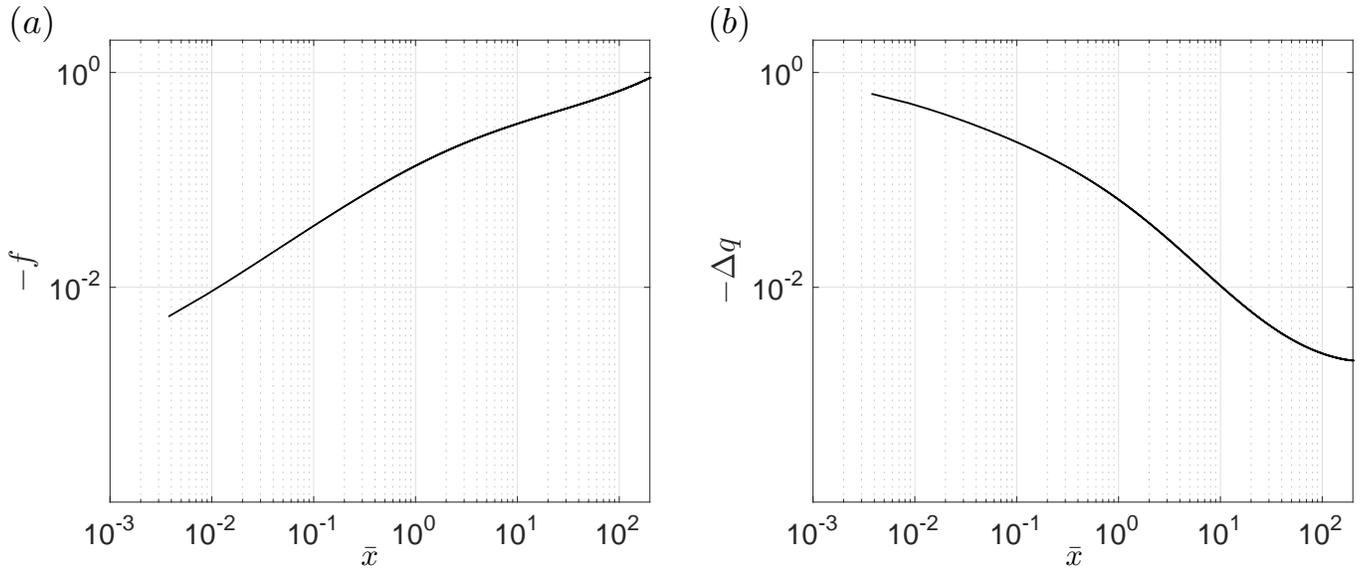}
	\caption{(a) The function $f(\bar{x})$, with $\bar{x}=\bar{z}-1$, such that $f(\bar{x}=0)=0$ and $f(\bar{x}\rightarrow \infty)\rightarrow -1$, is calculated solving the integral equation (\ref{Deltaq1}) using the numerical code provided in Appendix \ref{appB}. (b) The function $\Delta q(\bar{x})=df/d\bar{x}$ is such that $-\Delta q(\bar{x}\rightarrow 0)\rightarrow 0.7$.} 
   \label{fig3}
\end{figure*}

\begin{table}
\caption{\label{tab:1} The table provides the numerical values of the integral in Eq. (\ref{zjetslender}) for different values of $n$ using the values of $\Delta\,q(\bar{x})$ depicted in Fig \ref{fig3}.}
\vspace*{3mm}
\begin{ruledtabular}
\begin{tabular}{cccccccc}
   $\quad$ &  $n=2$ & $n=3$ & $n=4$ & $n=5$ & $n=6$ & $n=8$ & $n=10$ \\
   $1+n\,\int_0^\infty |\Delta\,q(\bar{x})|/(1+\bar{x})^{(n+1)} d\bar{x}$ & $1.14$ & $1.18$ & $1.20$ & $1.20$ & $1.24$ & $1.24$ & $1.30$
\end{tabular}
\end{ruledtabular}
\end{table}

In order to solve the integral equation for $\Delta\,q$, we first notice that, in terms of the variables 
\begin{equation}
\bar{z}=\frac{z}{z_{jet}(\tau)}\, ,\quad \bar{z}_0=\frac{z_0}{z_{jet}(\tau)}\, ,\quad \Delta\,q=\frac{df}{d\bar{z}}
\end{equation}
with $f(\bar{z}=1)=0$, the integral equation (\ref{Deltaq0b}) reads:
\begin{equation}
\begin{split}
&\frac{1}{\bar{z}-1}-\frac{1}{\bar{z}+1}+\int_{1}^\infty\frac{df/d\bar{z}_0\left(\bar{z}-\bar{z}_0\right)d\,\bar{z}_0}{\left[\left(\bar{z}-\bar{z}_0\right)^2\right]^{3/2}}=0\, .
\end{split}\label{Deltaq1}
\end{equation}

Equation (\ref{Deltaq1}) is solved numerically using the method detailed in Appendix \ref{appB}, and the resulting functions $f$ and $\Delta\,q$ are represented in Fig \ref{fig3}, where it is also shown that $-\Delta q(\bar{z}\rightarrow 1)\approx 0.7$. Let us point out here that the predicted sink distribution $q_\infty\left(1+|\Delta q|\right)$ with $\Delta\,q$ given in Fig. \ref{fig3} is compared in Fig. \ref{fig11} of Appendix \ref{appB} with the results of numerical simulations of the type described in Appendix \ref{appA}, finding good agreement between the theoretical and the numerical results. Once $\Delta\,q$ is known, the equation for $z_{jet}(\tau)$ is deduced following the same ideas as those in \cite{wagner1932} namely, making use of the kinematic boundary condition at the interface and considering that the jet root is located at the position where the interface meets the axis of symmetry:
\begin{equation}
\frac{dr_s}{d\tau}=v_r(r=r_s)=-q_{\infty}\frac{1+|\Delta\,q|}{r_s}\Rightarrow -\frac{r^2_s(z=z_{jet})}{2\,q_\infty}=-\tau-\int_0^{\tau} |\Delta\,q(z=z_{jet}(\tau))|\,d\tau'\, .\label{Wagner0}
\end{equation}
At an instant of time $\tau'\leq \tau$ and for a fixed value of $z_{jet}(\tau)$, we define
\begin{equation}
\bar{x}=z_{jet}(\tau)/z'_{jet}-1\quad \mathrm{with}\quad z'_{jet}=z_{jet}(\tau') \label{xdef}
\end{equation}
and, therefore, the integral equation (\ref{Wagner0}) can be written as
\begin{equation}
-\frac{r^2_s(z=z_{jet})}{2 q_\infty}=-\tau-\int_\infty^{0} |\Delta\,q(\bar{x})|\,\frac{d\tau'}{dz'_{jet}}\frac{dz'_{jet}}{d\bar{x}}\,d\bar{x}\, .\label{Wagner1}
\end{equation}

\begin{figure*}
	\begin{center}
	\includegraphics[width=\textwidth]{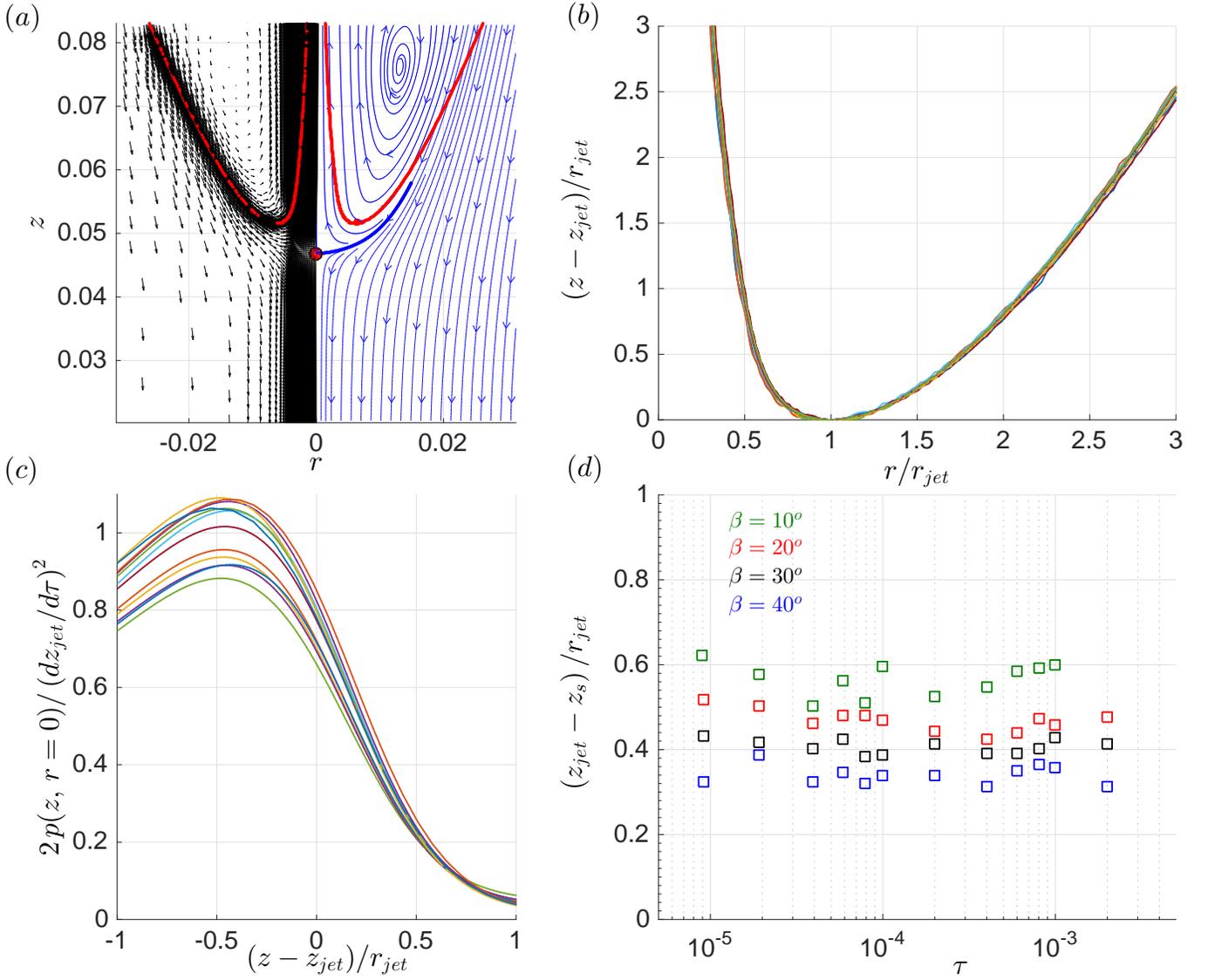}
	\end{center}
	\caption{$(a)$ Streamlines corresponding to a conical cavity with an opening semiangle of $\beta=20^\circ$, see Appendix \ref{appA} for numerical details, represented at a generic instant $\tau_0$ in a frame of reference moving vertically with the velocity $d z_{jet}/d\tau(\tau_0)$. The thick blue line represents the dividing streamline. $(b)$ When scaled using $r_{jet}(\tau)$, the jet shapes corresponding to $\beta=20^\circ$ superimpose onto a single curve for over two decades in time. $(c)$ Scaled pressure at the axis of symmetry in a region surrounding the base of the jet for over two decades in time; here, $\beta=20^\circ$. Pressures are scaled in terms of the value of the stagnation pressure in the moving frame of reference, $1/2\left(d z_{jet}/d\tau\right)^2$. $(d)$ Time evolution of the location of the stagnation point of the flow in the moving frame of reference, see the sketch in Fig \ref{fig1}(b). Green, red, black and blue squares indicate, respectively, the results corresponding to conical cavities with opening semiangles $\beta=10^\circ,\,20^\circ,\,30^\circ$ and $\beta=40^\circ$, see Appendix \ref{appA} for numerical details.} 
   \label{fig4}
\end{figure*}

Equation (\ref{Wagner1}) can be solved using a procedure similar to that followed in \cite{PRLSplash}, where we considered the analogous case of the impact of a spherical drop over a wall: indeed, for the case of axisymmetric cavities with equations of the type $r^2_s(z=z_{jet})=2C^{-1}\,z^n_{jet}$ with $C$ an arbitrary constant, the solution of equation (\ref{Wagner1}) is of the form $\tau=A^{-1}\,z^n_{jet}$ and, consequently,
\begin{equation}
\begin{split}
&-\frac{r^2_s(z=z_{jet})}{2 q_\infty}=-\tau-\int_\infty^{0} -|\Delta\,q(\bar{x})|\,nA^{-1}\left(z'_{jet}\right)^{(n-1)}\frac{z'^2_{jet}}{z_{jet}(\tau)}\,d\bar{x}\Rightarrow\\& z_{jet}=\left(1+n\int_0^\infty \frac{|\Delta\,q(\bar{x})|}{\left(1+\bar{x}\right)^{n+1}}\,d\bar{x}\right)^{1/n}\left(Cq_\infty\tau\right)^{1/n}\, ,\label{zjetslender}
\end{split}
\end{equation}
where we have made use of Eq. (\ref{xdef}). Notice that the values of the integral $\int_0^\infty |\Delta\,q|/(1+\bar{x})^{(n+1)} d\bar{x}$, which can be calculated numerically for different values of $n$ using the function $\Delta q(\bar{x})$ depicted in Fig \ref{fig3}, are given in table \ref{tab:1}. Once $z_{jet}(\tau)$ is calculated through equation (\ref{zjetslender}), the jet velocity in the laboratory frame of reference, $v_{jet}$, is calculated expressing the Euler-Bernoulli equation in a relative frame of reference moving with the jet base velocity, $d z_{jet}/d\tau$. Indeed, the numerical results in Fig \ref{fig4}(a), which correspond to the type of numerical simulations detailed in Appendix \ref{appA} indicate that, in the moving frame of reference, the free surface is a streamline along a vertical distance $\sim r_{jet}$ pointing upwards from the base of the jet and also that the stagnation point of the flow is located at a distance $\sim r_{jet}(\tau)$ below the base of the jet, see also Fig. \ref{fig1}(b). Moreover, Fig \ref{fig4}(a) reveals that the jet is fed from a narrow region of width $\sim r_{jet}(\tau)$ located nearby the free interface, a result which was already noticed by \cite{MacIntyre} experimentally and by \cite{PRL2009} numerically. In addition, Fig \ref{fig4}(b) shows that the jet shapes corresponding to different instants of time superimpose onto a single curve when rescaled using $r_{jet}(\tau)$ as the characteristic length scale, whereas Fig \ref{fig4}(c) shows the values of the rescaled pressure evaluated at the axis of symmetry along the spatial region $z-z_{jet}(\tau)\sim r_{jet}(\tau)$ for different instants of time. Figure \ref{fig4}(c) reveals that the stagnation point of the flow is indeed located at a distance $\sim r_{jet}$ below the base of the jet and also that the liquid pressure relaxes to that of the gas, $p_{gas}$, at $z\simeq z_{jet}(\tau)+r_{jet}(\tau)$. Therefore, since the jet is slender and capillary effects are subdominant, the pressure in the liquid will remain, in a first approach, constant for $z\gtrsim z_{jet}(\tau)+r_{jet}(\tau)$, this fact meaning that both the jet shape and the velocity within the spatio-temporal region $(z,\tau)$, with $z\gtrsim z_{jet}(\tau)+r_{jet}(\tau)$, can be predicted using the ballistic equations deduced in \cite{JFM2010}. Figure \ref{fig4}(c) also shows that the maximum pressure $p_{max}$ is the stagnation pressure, with this value being attained at the point where the dividing streamline meets the axis of symmetry, see Fig \ref{fig4}(a). Hence, $p_{max}$ can be calculated using the Euler-Bernoulli equation in the moving frame of reference as
\begin{equation} 
p_{max}-p_{gas}\simeq \frac{1}{2}\,\left(\frac{d z_{jet}}{d\tau}\right)^2\,  \label{pmax1} 
\end{equation}
because the modulus of the upstream velocity in the moving frame of reference at the dividing streamline is $\approx d z_{jet}/d\tau$ and the pressure upstream the dividing streamline, which is in close proximity to the free surface, is that of the gas because capillary effects are negligible. Therefore, taking into account that the flow field is quasisteady in a region of width $\sim r_{jet}(\tau)$ surrounding the base of the jet and making use of the Euler-Bernoulli equation along a streamline connecting the stagnation point of the flow in the relative frame of reference and the downstream position within the jet where the pressure relaxes to that of the gas, we find that 
\begin{equation} 
p_{max}-p_{gas}\simeq \frac{1}{2}\,\tilde{v}^2_{jet}\quad \mathrm{with}\quad \tilde{v}_{jet}=v_z(r=0,z=z_{jet}+r_{jet})-\frac{d z_{jet}}{d\tau}\,  \label{pmax2} 
\end{equation}
indicating the axial velocity in the moving frame of reference. Consequently, making use of equation (\ref{pmax1}) we conclude that the jet velocity in the laboratory frame of reference at the axial coordinate $z$ where the liquid pressure relaxes to that of the gas is given by
\begin{equation}
v_{jet}(\tau)=v_z(r=0,z=z_{jet}+r_{jet})=2\,\frac{d z_{jet}}{d\tau}\, .\label{vballistic}
\end{equation} 
Finally, the result in Fig \ref{fig4}(d) indicates that the maximum value of the pressure namely, the stagnation pressure in the moving frame of reference, see also the sketch in Fig. \ref{fig1}(b), is reached at a distance from the base of the jet $z_{jet}(\tau)-z_{s}(\tau)$, such that the ratio $\left(z_{jet}(\tau)-z_{s}(\tau)\right)/r_{jet}(\tau)\sim O(1)$ remains approximately constant in time.

 Then, making use of Eqs. (\ref{zjetslender}) and (\ref{vballistic}), we conclude that the velocity within the jet at the location where the liquid pressure relaxes to that of the gas, namely, at $z=z_{jet}(\tau)+\alpha\,r_{jet}(\tau)$ with $\alpha\simeq 1$, is:
\begin{equation}
\begin{split}
&v_{jet}\simeq 2\frac{d\,z_{jet}}{d\tau}\simeq\frac{2\,C\,q_\infty}{n}\left(1+n\int_0^\infty \frac{|\Delta\,q(\bar{x})|}{\left(1+\bar{x}\right)^{n+1}}\,d\bar{x}\right)^{1/n}\left(Cq_\infty\tau\right)^{(1-n)/n}\, .\label{vjetslender}
\end{split}
\end{equation}

The particularization of Eqs. (\ref{zjetslender}) and (\ref{vjetslender}) to the case of conical cavities with an equation of the form $r_s=z\tan\beta$, ($C=2/\tan^2 \beta$) yields 

\begin{equation}
z_{jet}=\left(1+2\int_0^\infty \frac{|\Delta\,q(\bar{x})|}{\left(1+\bar{x}\right)^{3}}\,d\bar{x}\right)^{1/2} 
\left(2\,q_\infty\,\tau/\tan^2 \beta\right)^{1/2} \simeq \frac{1.5}{\tan\beta}\sqrt{q_\infty\tau}\label{zjetconico}
\end{equation}
and
\begin{equation}
v_{jet}\simeq 2\frac{d z_{jet}}{d\tau}\simeq \frac{1.5}{\tan{\beta}}\sqrt{\frac{q_\infty}{\tau}}\, ,\label{vjetconico}
\end{equation}
where we have made use of the value of the integral given in table \ref{tab:1} corresponding to $n=2$. For the analogous case of parabolic cavities with an equation of the form $r_s=z^2/\left(2r_c\right)$, Eqs. (\ref{zjetslender}) and (\ref{vjetslender}) yield the following expressions for $z_{jet}$ and $v_{jet}$:
\begin{equation}
z_{jet}=1.25\left(2 r_c\right)^{1/2} \left(q_\infty\tau\right)^{1/4}\quad\mathrm{and}\quad v_{jet}=0.625\,q_\infty\,\left(2 r_c\right)^{1/2}\,\left(q_{\infty}\,\tau\right)^{-3/4} \, ,\label{zjetvjetparabolico}
\end{equation}
where we have made use of the value of the integral given in table \ref{tab:1} corresponding to $n=4$.

In order to deduce the equation for $r_{jet}(\tau)$, notice first that the sink distribution along the axis of symmetry $q_\infty\left(1+|\Delta\,q|\right)$, with $\Delta\,q$ the function represented in Fig \ref{fig3}, exactly balances the flow rate imposed as the far field boundary condition in Eq. (\ref{farfield}). However, as it is shown in Fig \ref{fig4}(a), the interface is a streamline in a frame of reference moving vertically with the velocity $d z_{jet}/d\tau$ along a distance $\sim r_{jet}$ extending upwards from the base of the jet, a fact meaning that the normal velocities to the interface in the moving frame of reference are zero in this region. Therefore, mass conservation enforces the emergence of a jet of width $\sim r_{jet}$ in order to compensate the flow rate which cannot be suctioned by the sink distribution $q_\infty\left(1+|\Delta\,q|\right)$ where the normal velocities are zero in the moving frame of reference. Consequently, mass conservation demands that the flow rate which cannot be suctioned by the interface through this region of length $\sim r_{jet}$, is expelled into the gaseous atmosphere in the form of a jet of width $r_{jet}$ given by:
\begin{equation}
2\pi\int_{z_{jet}}^{z_{jet}+r_{jet}}q_\infty\left(1+|\Delta\,q(z)|\right) dz\approx 2\pi r^2_{jet}\frac{d z_{jet}}{d\tau} \quad \Rightarrow \quad r_{jet}\frac{d z_{jet}}{d\tau}\simeq 1.7 q_\infty\, ,\label{balrjet}
\end{equation}
where we have taken into account that: i) the rate of volume increase in the bulk of the liquid is $\pi r^2_{jet} d z_{jet}/d\tau$, ii) by virtue of the Euler-Bernoulli equation, the jet velocity in the moving frame of reference is $d z_{jet}/d\tau$ and iii) the function $\Delta q$, represented in Fig. (\ref{fig10}) is such that $-\Delta q(\bar{x}\rightarrow 0)\rightarrow 0.7$ and, consequently,
\begin{equation}
\int_{z_{jet}}^{z_{jet}+r_{jet}}\,|\Delta\,q|\,dz \simeq 
|\Delta q(\bar{x}=0)|\,r_{jet}\approx 0.7 r_{jet}\, .
\end{equation}
\begin{figure*}
	\centering
	\includegraphics[width=0.9\textwidth]{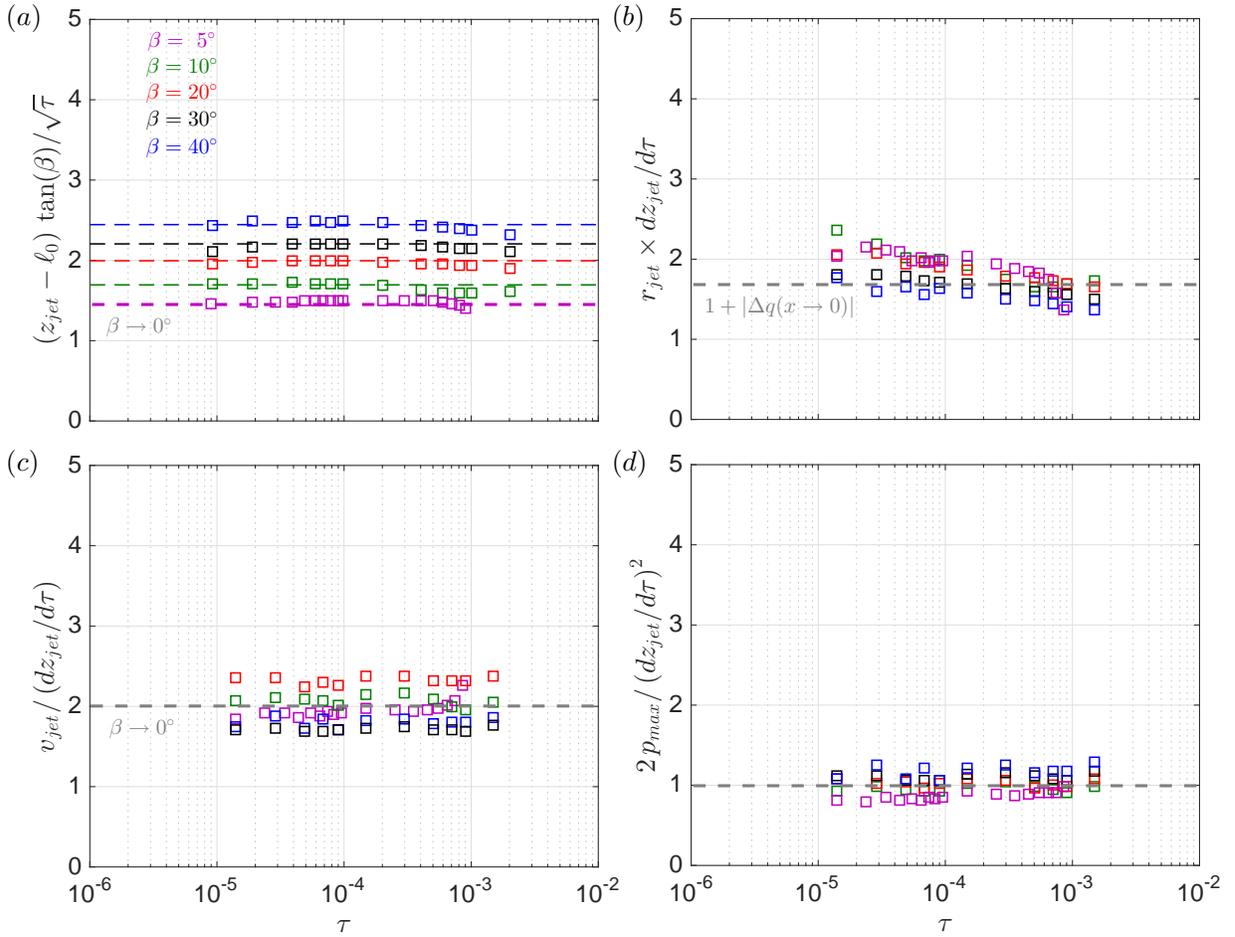}
	\caption{Comparison between the numerical results in Appendix \ref{appA} corresponding to the case of conical cavities with different values of the opening semiangle $\beta$ with the predictions in Eqs. (\ref{pmax1}), (\ref{zjetconico}), (\ref{vjetconico}) and (\ref{balrjet}). Here, $v_{jet}(\tau)=v_z(z=z_{jet}(\tau)+\alpha\,r_{jet}(\tau))$ with $\alpha=1$ for the case of slender cavities namely, the conical cavities with $\beta=5^\circ$, $\beta=10^\circ$ and $\beta=20^\circ$, see Fig. \ref{fig4}(c), whereas $\alpha=0$ for the cases of non slender conical cavities namely, those with $\beta=30^\circ$ and $\beta=40^\circ$. The reason for the different values of $\alpha$ is that the liquid pressure relaxes to that of the gas at a different vertical locations within the jet depending on the value of the opening semiangle, see Fig. \ref{fig4}(c) for the case $\beta=20^\circ$. Here, $\ell_0(\beta=5^\circ)=0.0080$, $\ell_0(\beta=10^\circ)=0.0045$, $\ell_0(\beta=20^\circ)=0.0028$, $\ell_0(\beta=30^\circ)=0.0022$, $\ell_0(\beta=40^\circ)=0.0012$.} 
   \label{fig5}
\end{figure*}

 In order to validate the results in Eqs. (\ref{zjetslender}), (\ref{pmax1}), (\ref{vjetslender}),  and (\ref{balrjet}) we make use of the results of the numerical simulations described in Appendix \ref{appA}. Indeed, in  Appendix \ref{appA} we analyze the results of numerical simulations using the \texttt{GERRIS} \citep{Popinet2003,Popinet2009} script used by \cite{Reuter} to simulate the emergence of very fast jets, with velocities up to $1000$ $m\,s^{-1}$, which are emitted when a bubble cavitates in very close proximity to a solid wall once the value of the flow rate per unit length is prescribed as a far field boundary condition. At this point, notice that our own analysis of the governing equations and of the numerical results reported in Appendix \ref{appA} permits us to conclude that both the jet shapes and the velocity field converge towards a $\beta$-dependent self-similar solution of the Euler-Bernoulli and Laplace equations such that lengths and velocities are respectively proportional to $\sqrt{q_\infty\tau}$ and $\sqrt{q_\infty/\tau}$ for arbitrary values of $\beta$ a result which, up to our knowledge, had not been reported before. 

Figure \ref{fig5} compares the predictions given in Eqs. (\ref{pmax1}), (\ref{zjetconico}), (\ref{vjetconico}) and (\ref{balrjet}) with the numerical values of $r_{jet}(\tau)$, $v_{jet}(\tau)$, $p_{max}(\tau)$ and of $r_{jet}d z_{jet}/d\tau(\tau)$ obtained from the numerical simulations detailed in Appendix \ref{appA} for the case of conical cavities with different values of the opening semiangle $\beta$ taking $q_\infty=1$. The results depicted in Fig. \ref{fig5} confirm our predictions in the limit $\beta\ll 1$ and also permit us to extend our results to generic values of $\beta$ by just writing 
\begin{equation}
z_{jet}(\beta,\tau)=K(\beta)z_{jet}(\beta\ll 1,\tau)=K(\beta)\frac{1.5}{\tan\beta}\sqrt{q_\infty\tau}\, \quad \mathrm{with}\quad K(\beta\leq 5^\circ)=1\quad\mathrm{and}\label{zjetconico2}
\end{equation}
$K(\beta=10^\circ)=1.13$, $K(\beta=20^\circ)=1.33$, $K(\beta=30^\circ)=1.47$ and $K(\beta=40^\circ)=1.63$. 

\begin{figure*}
	\centering
	\includegraphics[width=0.9\textwidth]{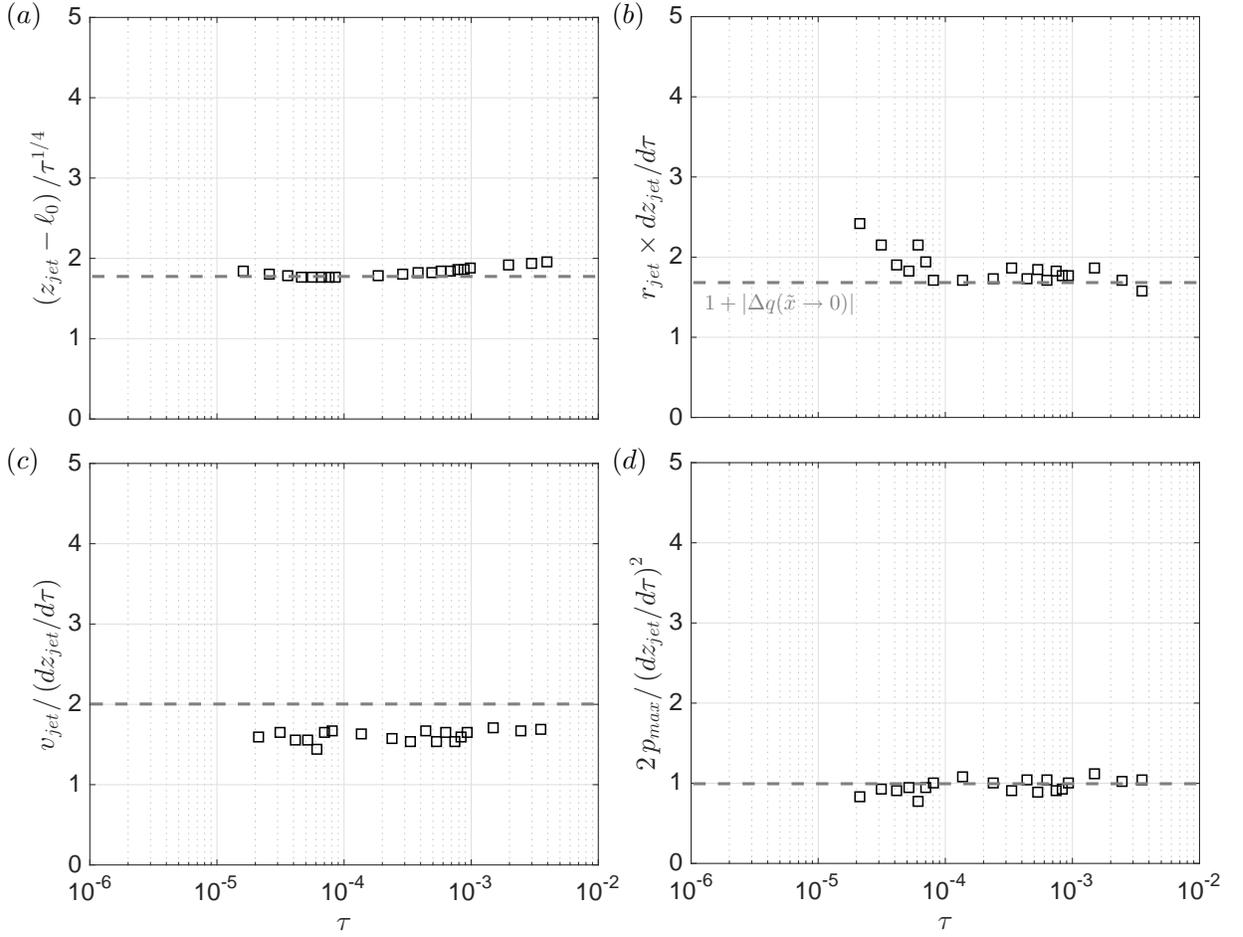}
	\caption{Comparison between the numerical results of $p_{max}(\tau)-p_{gas}$, $z_{jet}(\tau)$, $v_{jet}(\tau)=v_z(r=0,z=z_{jet}+r_{jet})$, $r_{jet}(\tau)$ and the respective values predicted by equations (\ref{pmax1}), (\ref{zjetvjetparabolico}) and (\ref{balrjet}). The reason for the small deviation between the predicted and the calculated jet velocity is because the liquid pressure relaxes to that of the gas at a vertical position within the jet which is slightly above $z=z_{jet}(\tau)+r_{jet}(\tau)$. Here, $\ell_0(r_c=1)=0.015$.} 
   \label{fig6}
\end{figure*}

Moreover, Fig. \ref{fig6} compares the predictions in Eqs. (\ref{pmax1}), (\ref{zjetvjetparabolico}) and (\ref{balrjet}) with the numerical values of $r_{jet}(\tau)$, $v_{jet}(\tau)$, $p_{max}(\tau)$ and of $r_{jet}d z_{jet}/d\tau(\tau)$ calculated using the same type of simulations as those detailed in Appendix \ref{appA} but replacing the equation for the initial shape of the interface by a parabola of equation $r_s(z,\tau=0)=z^2/2$. The results shown in Fig. \ref{fig6}, which do not include any type of adjustable constant, provide further support to our theory.

\section{Conclusions}
\label{sec:Con}

To conclude, here we have analized the high-Reynolds number implosion of cavities when the collapse is driven by a far field boundary condition expressing that the flow rate per unit length remains constant in time. We have presented a self-consistent theory which describes the collapse of slender cavities with a generic geometry and provides with algebraic equations, with no adjustable constants, for the time evolutions of the speed, width and of the vertical position of the jets. When particularized to the case of conical bubbles, our algebraic equations reproduce the self-similar results obtained numerically in the limit of small values of the opening semiangle and, when particularized to the case of parabolic cavities, we also find that our predictions for the jet radius, jet velocity and the maximum liquid pressure are in good agreement with simulations. We have shown both in this contribution and also in \cite{PRL2023} that the jets produced right after a bubble bursts at an interface can be described in terms of our purely inertial theory when the jet emerges from the base of a truncated cone, a result which differs from the common belief that the dynamics of the jets produced after the inertio-capillary collapse of cavities can be described in terms of an inertio-capillary balance see e.g., \cite{PRLEggers}. The results presented here and in \cite{PRL2023} might also be applied to describe the type of very high speed jets generated when a conical bubble implodes very close to a wall, a phenomenon which could play a role in the degradation of the material produced by liquid cavitation.
 
\begin{acknowledgements}
This research work has been partially supported by the Grant PID2020-115655G financed by the Spanish MCIN/ AEI/10.13039/501100011033.\\

\emph{Author contributions:} JMG designed the research and the theory, analyzed the data and wrote the paper, FJBR performed the numerical simulations and analyzed the data. All authors reviewed the results and approved the final version of the manuscript.
\end{acknowledgements}


\appendix
\section{Self-similar conical jets}
\label{appA}

\begin{figure*}
	\centering
	\includegraphics[width=\textwidth]{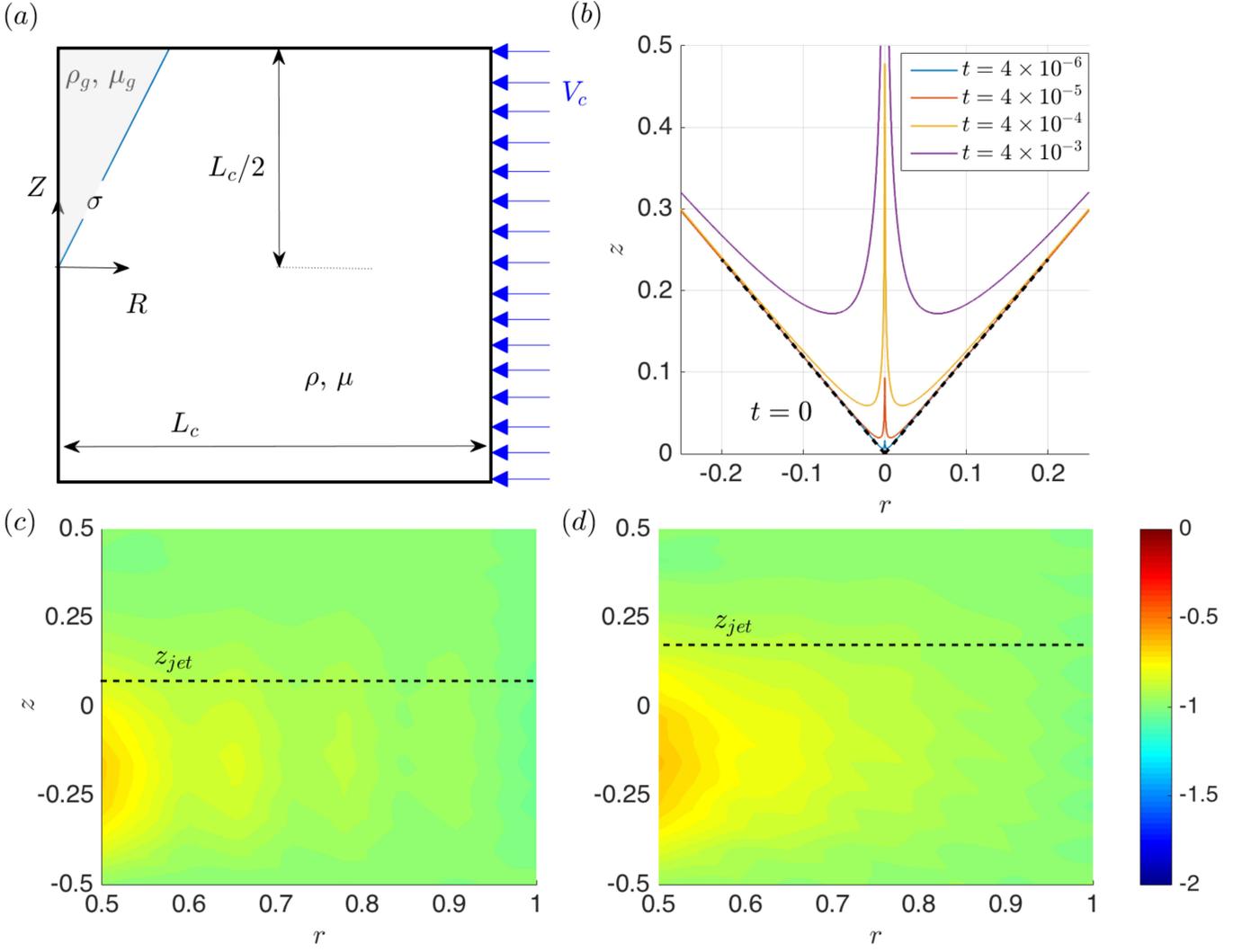}
	\caption{(a) Sketch of the numerical setup indicating the far field boundary condition, the initial shape of the conical interface, the material properties of the liquid and the gas as well as the radial and axial coordinates, $R$ and $Z$ respectively. (b) Numerical results corresponding to $Re=100$ and $We=1$ and $\beta=40^\circ$. (c)-(d) Numerical values of $r v_r$ corresponding to the simulations depicted in (b) at two different instants of time: $\tau=4\times 10^{-4}$ (c) and $\tau=4\times 10^{-3}$ (d). Notice that the type of numerical simulations carried out using the boundary conditions sketched in (a) impose, as a far field boundary condition, a constant value of the flow rate per unit length, $r v_r\rightarrow -1$.} 
   \label{fig7}
\end{figure*}

In this Appendix we present numerical results describing the high-Reynolds number implosion of conical voids such as those found in the collapsing depressions of standing waves \citep{Zeff}, during the collapse of bursting bubbles \cite{Duchemin,JFM2019} or as a consequence of the collapse of the cavity formed after a drop impacts a liquid pool \citep{Michon,Thoroddsen_2018,JFM2021}. The numerical results have been obtained using the open-source package
\texttt{GERRIS} \citep{Popinet2003,Popinet2009} and therefore, the effects of compressibility on the collapse and jet ejection processes are not retained in the simulations. In fact, the results in this section have been obtained using the \texttt{GERRIS} script provided by \cite{Reuter} in their study of the dynamic of jets produced in cavitation bubbles near a rigid boundary for small values of the stand-off parameter. The numerical setup in Fig \ref{fig7}(a) illustrates the radial and axial coordinates, $R$ and $Z$ respectively, the initial shape of a conical cavity with a half-opening angle $\beta$ and also $L_c$, namely, the radius of the cylindrical surface where the value of the far-field radial velocity, $V_c$, is imposed. In the following, $\rho$, $\mu$ and $\sigma$ will respectively denote the liquid density, viscosity and interfacial tension coefficient, whereas $\rho_g$ and $\mu_g$ will be used to refer to the gas density and viscosity. The numerical code is solved fixing the value of the flow rate per unit length, $L_c V_c$, at $R=L_c$ and imposing outflow  boundary conditions at $Z=\pm L_c/2$  i.e., the  $Z$-derivaties of the velocity and pressure fields are zero at the top and bottom boundaries. The minimum grid size is $2^{-12} L_c$ and the jet radius at the region where it meets the conical surface, see Fig \ref{fig7}(b), will be resolved with at least $6$ numerical cells.

\begin{figure*}
	\centering
	\includegraphics[width=\textwidth]{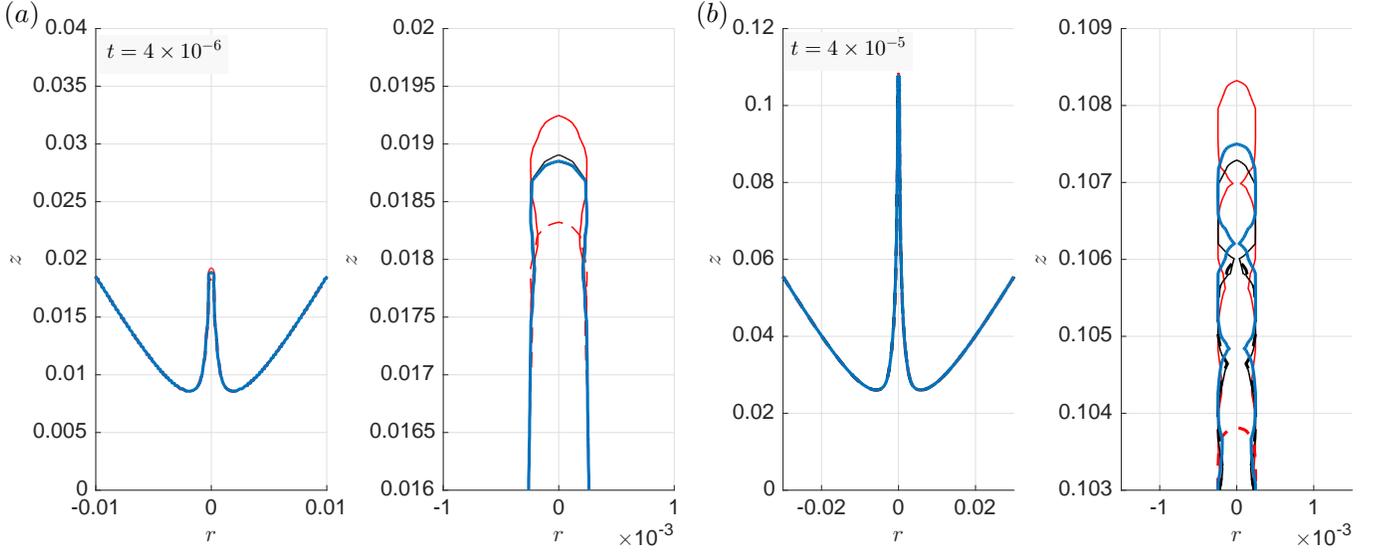}
	\caption{Analysis of the influence of $Re$ and $We$ on the jet ejection process for a fixed value of $\beta=30^\circ$ and two different instants of time $\tau=4\times 10^{-6}$ (a) and $\tau=4\times 10^{-5}$ (b). The results in the figure have been obtained for the following values of the control parameters: $Re=100$, $We=1$ (blue continuous line), $Re=50$, $We=1$ (red dashed line), $Re=200$, $We=1$ (red continuous line), $Re=100$, $We=5$ (black continuous line), $Re=100$, $We=0.5$ (black dashed line). Notice that the differences between the jet shapes are practically indistinguishable and are only appreciable in the zoomed images corresponding to the jet tip, a fact indicating that the spatio-temporal evolution of the jet is virtually independent of $Re$ and $We$.} 
   \label{fig8}
\end{figure*}
\begin{figure*}
	\centering
	\includegraphics[width=\textwidth]{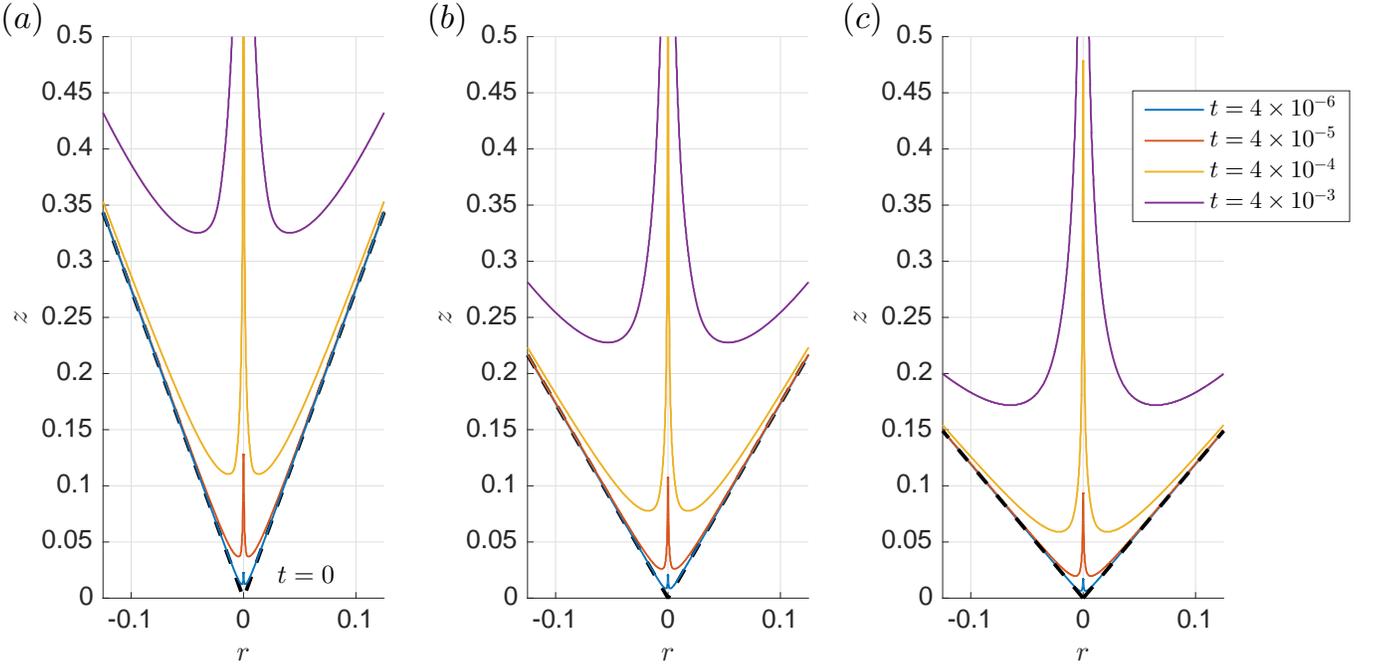}
	\caption{This figure analyzes the influence of the opening semiangle on the spatio-temporal evolution of the jet: (a) $\beta=20^\circ$, (b) $\beta=30^\circ$, (c) $\beta=40^\circ$. In the three cases considered, $Re=100$ and $We=1$. Notice that the jet speed increases when $\beta$ decreases.} 
   \label{fig9}
\end{figure*}

Using $L_c$, $L_c/V_c$ and $\rho V^2_c$ as the characteristic scales of length, time and pressure, the flow sketched in Fig \ref{fig7} can be described in terms of the following dimensionless parameters:
\begin{equation}
We=\frac{\rho\,V^2_c\,L_c}{\sigma}\, ,\quad Re=\frac{\rho\,V_c L_c}{\mu}\, ,\quad m=\frac{\mu_g}{\mu}\, ,\quad \Lambda=\frac{\rho_g}{\rho}\quad\mathrm{and}\quad \beta \label{WeRe1}
\end{equation}
but, since here we fix the values of the density and viscosity ratios to those characterizing the air-water system, $\Lambda=1.2\times 10^{-3}$, $m=1.8\times 10^{-2}$, the numerical solutions will only depend on $We$, $Re$ and $\beta$. Notice also that, from now on, dimensionless variables will be written using lower case letters to differentiate them from their dimensional counterparts, written in capitals, and $\tau$ and $\mathbf{n}$ will indicate, respectively, the dimensionless time after the jet is ejected and the unit normal vector to the interface of equation
\begin{equation}
r=r_s(z,\tau)\Rightarrow F=r-r_s(z,\tau)=0\, .\label{Kinematic}
\end{equation}

Figure \ref{fig8} shows that the numerical results are virtually independent of both $We$ and $Re$ for the range of values of these two parameters considered in this study namely, $Re\geq 50$ and $We\geq 0.5$. While the results regarding the effect of the Reynolds number could have been anticipated in view of the fact that the minimum value of $Re$ is already large, $Re=50$, the independence of the solution with $We$ indicates that the mechanism of jet ejection is not driven by capillarity. The influence of the opening semiangle $\beta$ on the temporal evolution of the jet shapes is analyzed in Fig \ref{fig9}, where it can be clearly appreciated that the jet velocity increases when $\beta$ decreases.

It is our purpose now to explore whether there exists a self-similar solution for the high-Reynolds number jets depicted in Fig \ref{fig9} for instants of time such that $\tau\ll 1$. The analysis starts by considering that the vorticity production at the interface can be neglected in the description of the jet dynamics in the limit $Re\gg 1$ and then, the velocity field $\mathbf{v}$ can be expressed in terms of the velocity potential $\phi$ as
\begin{equation}
\mathbf{v}=\nabla\phi\, .\label{virr}
\end{equation} 
Therefore, in the incompressible limit of interest here, continuity demands that
\begin{equation}
\nabla\cdot \mathbf{v}=0\Rightarrow \nabla^2\phi=0\, ,\label{Laplace}
\end{equation}
where we have made use of equation (\ref{virr}). The Laplace equation (\ref{Laplace}) must be solved subjected to the value of the velocity potential calculated particularizing the Euler-Bernoulli equation 
\begin{equation}
\frac{\partial\phi}{\partial \tau}+\frac{|\nabla\phi|^2}{2}-We^{-1}\nabla\cdot\mathbf{n}=C(\tau)\label{E-B}
\end{equation}
at the interface of equation $F=0$, with $F$ the function defined in (\ref{Kinematic}) satisfying the kinematic boundary condition,
\begin{equation}
\frac{DF}{Dt}=0\Rightarrow \frac{\partial \phi}{\partial r}=\frac{\partial r_s}{\partial \tau}+\frac{\partial \phi}{\partial z} \frac{\partial r_s}{\partial z}\, ,\quad\mathrm{with}\quad r_s(z\rightarrow\infty,\tau)\rightarrow z\tan\beta\label{F-S}
\end{equation}
and with $D/Dt$ indicating the material derivative. In \cite{Zeff}, the self-similar structure of the flow field was deduced introducing the scaled variables
\begin{equation}
\phi=\left(\tau-\tau_0\right)^\delta\,h\left[\frac{z-\ell_0}{\left(\tau-\tau_0\right)^\varepsilon},\frac{r}{\left(\tau-\tau_0\right)^\varepsilon}\right] \quad\mathrm{and}\quad
r_s(z,\tau)=\left(\tau-\tau_0\right)^\varepsilon\,g\left[\frac{z-\ell_0}{\left(\tau-\tau_0\right)^\varepsilon}\right]\, \label{ansatz}
\end{equation}
into the system of equations (\ref{Laplace})-(\ref{F-S}), finding that both equations (\ref{Laplace}) and (\ref{F-S}) can be written in terms of the self-similar variables defined in (\ref{ansatz}) if $\delta=2\varepsilon-1$, with this condition also ensuring that the terms $\partial \phi/\partial \tau$ and $|\nabla\phi|^2/2$  in the Euler-Bernoulli equation (\ref{E-B}) possess an identical functional dependence with time. In \cite{Zeff}, the value of the exponent $\varepsilon=2/3$ characterizing the length scale in their inertio-capillary self-similar solutions was deduced imposing the additional condition that the time dependence of the three terms at the left hand side of equation (\ref{E-B}) must be the same. 

However, the results depicted in Fig \ref{fig8} reveal that the time evolution of the jet is independent of the Weber number, a fact indicating that the type of jets considered here are not forced by capillarity. Consequently, in the present case, the only restriction imposed by equations (\ref{Laplace})-(\ref{F-S}) for self-similar solutions to exist is that $\delta=2\varepsilon-1$ and then, the value of the exponent $\varepsilon$ in equation (\ref{ansatz}) should be fixed by the far field boundary condition of the Laplace equation (\ref{Laplace}) which, in our case, reads -see Fig. \ref{fig7}(c)-(d):
\begin{equation}
r\frac{\partial\phi}{\partial r}\rightarrow -1\quad\mathrm{for}\quad r\rightarrow\infty\, .\label{qinfty}
\end{equation}
The far field boundary condition (\ref{qinfty}) differs from the one corresponding to the inertio-capillary collapse of a cavity, which happens to be analogous to that driving the self-similar wave created after the impact of a disc over a liquid pool. Indeed, in this latter case, \cite{Iafrati,JFMIvo} reported that the far-field boundary condition for the velocity potential is the well-known two-dimensional irrotational flow around a wedge of angle $2\pi$,
\begin{equation}
\phi(r\rightarrow\infty)\propto r^{1/2}w(\theta)\, ,\label{BCIvo}
\end{equation}
with $r$ and $\theta$ indicating the polar coordinates. The different boundary conditions (\ref{BCIvo}) and (\ref{qinfty}) explain the differences between the exponent $\varepsilon=2/3$ characterizing the self-similar solutions reported by \cite{Zeff,JFMIvo} and the corresponding value of $\varepsilon$ which is imposed by equation (\ref{qinfty}). Indeed, here $\varepsilon=1/2$  because, as it was pointed out above, $\delta=2\varepsilon-1$ and, in addition, equation (\ref{qinfty}) can only be written in terms of the self-similar variables defined in (\ref{ansatz}) if $\delta=0$. Since $\varepsilon=1/2$, in the case the numerical solution converged to a self-similar solution of the system of equations and boundary conditions for sufficiently large values of $\tau>\tau_0$, the characteristic length scale of the jets depicted in Fig. \ref{fig8} should depend on time as $\left(\tau-\tau_0\right)^{1/2}$ whereas velocities should depend on time as $\left(\tau-\tau_0\right)^{-1/2}$, with these predictions also implying that, in the limit $\left(\tau-\tau_0\right)\ll 1$ of interest here, both $C$ and also the capillary term in equation (\ref{E-B}) are subdominant with respect to the inertial terms $\partial \phi/\partial \tau$ and $|\nabla\phi|^2/2$. 

\begin{figure*}
	\centering
	\includegraphics[width=0.6\textwidth]{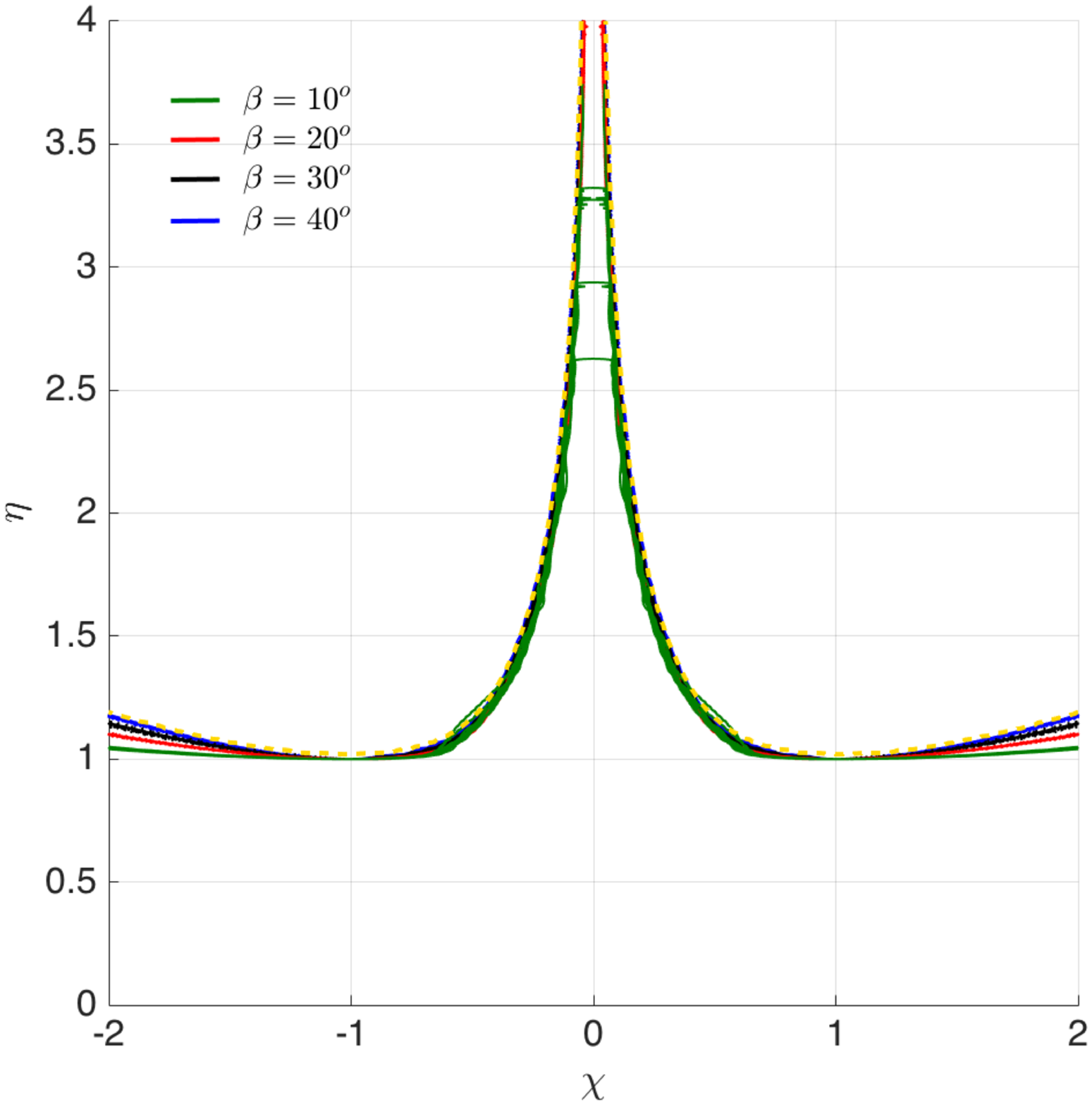}
	\caption{The spatio-temporal evolution of the jet shapes of the type shown in Fig \ref{fig9} collapse, along two decades in time and for four different values of $\beta$ namely, $\beta=10^\circ,\, 20^\circ,\,30^\circ$ and $\beta=40^\circ$, onto the $\tau$-independent and almost $\beta$-independent function depicted in the figure, with $\chi$ and $\eta$ defined in equation (\ref{varsem}). This function for the self-similar shapes of the jets is compared in the Supplementary Material of \cite{PRL2023} with the numerical results obtained for the case of bubble bursting jets emerging from the base of a truncated conical surface.} 
   \label{fig10}
\end{figure*}

In view of the previous analysis, here explore whether the different jet shapes depicted in Fig. \ref{fig9} can be collapsed onto a single curve. For this purpose, we define the time-dependent functions $z_{jet}(\tau,\beta)$ and $r_{jet}(\tau,\beta)$ illustrated in Fig \ref{fig1}. Figure \ref{fig10} shows that the time-dependent jet shapes depicted in Fig \ref{fig9} do indeed collapse for over two decades in time onto a time-independent and almost $\beta$-independent function when plotted in terms of the scaled radial and vertical coordinates suggested by our previous analysis and defined as
\begin{equation}
\chi=\frac{r}{r_{jet}(\tau,\beta)}\, ,\quad \mathrm{and} \quad\eta=\frac{z-\ell_0(\beta)}{z_{jet}(\tau,\beta)-\ell_0(\beta)}\ ,\label{varsem}
\end{equation}
with the values of $\ell_0(\beta)$ given in Fig. \ref{fig5}, where it is depicted that, indeed, $z_{jet}(\tau,\beta)-\ell_0(\beta)\propto \tau^{1/2}$, $d\,z_{jet}/d\tau\propto \tau^{-1/2}$ and $r_{jet}(\tau,\beta)\propto \tau^{1/2}$.

The numerical results corresponding the case of a parabolic cavity  shown in the main text in Fig. \ref{fig6}, have been obtained using the same numerical code and the same boundary conditions as those detailed in this Appendix, being the only difference that the equation for the initial shape of the interface is, in this case, $r_s(z,\tau=0)=z^2/2$.  

\section{Solution of the integral equation for $\Delta\,q$}
\label{appB}

Equation (\ref{Deltaq1}) is solved integrating by parts, once we notice that the integrands in Eq. (\ref{Deltaq1}) can be written, for $\bar{z}>\bar{z}_0$, as:
\begin{equation}
\frac{df/d\bar{z}_0}{\left(\bar{z}-\bar{z}_0\right)^2}=\frac{d}{d\bar{z}_0}\left(\frac{f}{\left(\bar{z}-\bar{z}_0\right)^2}\right)-2\frac{f}{\left(\bar{z}-\bar{z}_0\right)^3}\label{parts}
\end{equation}
and for $\bar{z}<\bar{z}_0$ as:
\begin{equation}
-\frac{df/d\bar{z}_0}{\left(\bar{z}-\bar{z}_0\right)^2}=-\left[\frac{d}{d\bar{z}_0}\left(\frac{f}{\left(\bar{z}-\bar{z}_0\right)^2}\right)-2\frac{f}{\left(\bar{z}-\bar{z}_0\right)^3}\right]\, .
\end{equation}
Notice that
\begin{equation}
\lim_{\gamma\rightarrow 0}\int_{1}^{\bar{z}-\gamma}\frac{d}{d\bar{z}_0}\left(\frac{f}{\left(\bar{z}-\bar{z}_0\right)^2}\right)\,d\bar{z}_0=\frac{f(\bar{z})}{\gamma^2}
\end{equation}
where we have taken into account that $f(\bar{z}=1)=0$. Moreover, notice that
\begin{equation}
\lim_{\gamma\rightarrow 0}\int_{\bar{z}+\gamma}^{\infty} -\frac{d}{d\bar{z}_0}\left(\frac{f}{\left(\bar{z}-\bar{z}_0\right)^2}\right)d\bar{z}_0=\frac{f(\bar{z})}{\gamma^2}\, .
\end{equation}

\begin{figure*}
	\centering
	\includegraphics[width=0.6\textwidth]{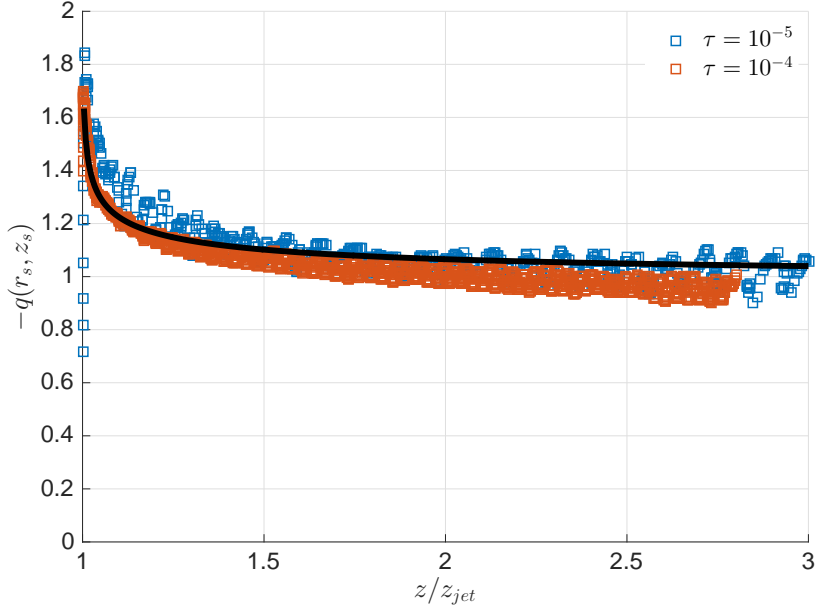}
	\caption{The numerical values of $-q(r_s,v_s)=-r_s(z,\tau) v_r(r=r_s,z,\tau)$ calculated at two instants of time for a value of the opening semiangle $\beta=5^\circ$ are compared with $q_\infty\left(1+|\Delta\,q|\right)$ with $\Delta\,q$ given in Fig. \ref{fig3} (black continuous line), finding good agreement between simulations and the theoretical prediction.} 
   \label{fig11}
\end{figure*}

We now discretise $f$, which is assumed to be constant and of value $f(\bar{z}_i)=f_i$ along $N$ panels of constant width $h$, centered at 
$\bar{z}_i=1+3h/4+\left(i-1\right)h$. Then, for $i>j$:
\begin{equation}
\int_{\bar{z}_j-h/2}^{\bar{z}_j+h/2}-\frac{2 f_j}{\left(\bar{z}_i-\bar{z}_0\right)^3}d\bar{z}_0=-f_j\left[\frac{1}{\left(\bar{z}_i-\bar{z}_j-h/2\right)^2}-\frac{1}{\left(\bar{z}_i-\bar{z}_j+h/2\right)^2}\right]\, ,
\end{equation}
for $i<j$:
\begin{equation}
\int_{\bar{z}_j-h/2}^{\bar{z}_j+h/2} \frac{2 f_j}{\left(\bar{z}_i-\bar{z}_0\right)^3}d\bar{z}_0=f_j\left[\frac{1}{\left(\bar{z}_i-\bar{z}_j-h/2\right)^2}-\frac{1}{\left(\bar{z}_i-\bar{z}_j+h/2\right)^2}\right]\, ,
\end{equation}
whereas, for $i=j$:
\begin{equation}
\begin{split}
&\int_{\bar{z}_i-h/2}^{\bar{z}_i-\gamma}-\frac{2 f_i}{\left(\bar{z}_i-\bar{z}_0\right)^3}d\bar{z}_0=-f_i\left[\frac{1}{\gamma^2}-\frac{1}{\left(h/2\right)^2}\right]\, \quad\mathrm{and}\\& 
\int_{\bar{z}_i+\gamma}^{\bar{z}_i+h/2} \frac{2 f_i}{\left(\bar{z}_i-\bar{z}_0\right)^3}d\bar{z}_0=f_i\left[-\frac{1}{\gamma^2}+\frac{1}{\left(h/2\right)^2}\right]\label{ij}
\end{split}
\end{equation}

For values $\bar{z}>\bar{z}_N+h/2$, with $\bar{z}_N\gg 1$ we impose mass conservation namely, that the flow rate injected by the distribution of sources located at $0\leq \bar{z}\leq 1$ equals the flow rate through the free interface, see Eq. (\ref{Deltaq1}):
\begin{equation}
\int_{1}^\infty \Delta q d\bar{z}=\int_{1}^\infty \frac{df}{d\bar{z}} d\bar{z}=f_\infty=-1
\end{equation}
and, therefore, the integral in Eq. (\ref{Deltaq1}) for $\bar{z}>\bar{z}_N+h/2$ can be solved analytically to give:
\begin{equation}
 \int_{\bar{z}_N+h/2}^\infty\frac{2f_\infty}{\left(\bar{z}_i-\bar{z}_0\right)^3}\,d\bar{z}_0=\frac{1}{\left(\bar{z}_i-\bar{z}_N-h/2\right)^2}\label{Intinfty}
\end{equation}
The numerical code provided as \textit{Supplementary Material} solves the integral equation (\ref{Deltaq1}) using the results in Eqs. (\ref{parts})-(\ref{Intinfty}) and, once the function $f$ is calculated solving the resulting linear system of equations, the function $\Delta\,q$ is calculated as
\begin{equation}
\Delta\,q_i=\Delta\,q(\bar{z}_i)=\frac{f_{i+1}-f_i}{h}\, .
\end{equation}

The results in Fig. \ref{fig11} compare the values of the flow rate at the interface calculated numerically as $-q(r_s,v_s)=-r_s(z,\tau) v_r(r=r_s,z,\tau)$ for a value of the opening semiangle $\beta=5^\circ$ with our prediction, $q_\infty\left(1+|\Delta\,q|\right)$, at two different instants of time, finding good agreement between our theory and the numerical results.




\end{document}